# Weak ferrimagnets of the YFe$_{1-x}$Cr$_x$O$_3$ type: negative magnetization and spin reorientation

EV Vasinovich [1], AS Moskvin [1,2]

[1] Ural Federal University, Ekaterinburg, Russia

[2] MN Mikheev Institute of Metal Physics, Ural Branch, Russian Academy of Sciences, Ekaterinburg, Russia

Abstract

In this work, we present an analysis of magnetic properties of weak ferrimagnets of the YFe$_{1-x}$Cr$_x$O$_3$ type. Taking into account the main spin interactions – isotropic Heisenberg superexchange, antisymmetric Dzyaloshinskii-Moriya exchange, single-ion spin anisotropy – calculations of the free energy, concentration and temperature dependences of magnetization were carried out within a framework of a molecular field model. In particular, the model demonstrates the phenomena of temperature compensation and negative magnetization, as well as spin reorientation. The compensation temperature reaches room temperature $T = 300$K in a composition with $x \approx 0.45$. The existence of magnetic structures of the $G_{xyz}$ type with the spatial orientation of the Néel vector is predicted.

## 1 Introduction

Rare earth orthoferrites-orthochromites of the type RFe$_{1-x}$Cr$_x$O$_3$ (R = Nd, Gd, Dy, Y, Lu) have been the subject of intensive fundamental theoretical and experimental studies in the second half of the 20th century due to the combination of their unique magnetic, magneto-optical, magnetoelastic and magnetic resonance properties, primarily weak ferro-and antiferromagnetism, spin-reorientation transitions (SR), the phenomenon of compensation of the magnetic moment and negative magnetization [1]. Moreover, studies of the mixed orthoferrite-orthochromite $YFe_{1-x}Cr_xO_3$, carried out more than 50 years ago, led to the theoretical prediction and experimental discovery of a new phenomenon - weak ferrimagnetism arising from the competition of the signs of the Dzyaloshinskii vector [2].

A new surge of interest in these systems already in the 21st century (see, for example, [3-12]) is associated with the opened prospects for the practical use of the phenomenon of temperature compensation of the magnetic moment and the related effects of "negative" magnetization and exchange bias for the creation of various multifunctional spintronic devices. However, in most of the "new" works devoted to the study of weak ferrimagnets of the $YFe_{1-x}Cr_xO_3$ type, we encounter both the ambiguity of the experimental data and the ambiguity in their interpretation.

In particular, regarding the Neel temperature $T_N$, it is only known that it decreases monotonically with a change in composition from $YFeO_3$ to $YCrO_3$ [2,10], but there is no consensus regarding the exact value of $T_N$ and, for example, values $T_N$ from $210 - 250$ K [3-7] and up to $T_N = 340 - 360$ K [2] for the composition $YFe_{0.5}Cr_{0.5}O_3$ are reported, which is apparently associated with the features of polycrystalline samples and small magnetization values. There are even fewer data on the compensation temperature $T_{comp}$ and they are also contradictory, for example, for the $YFe_{0.5}Cr_{0.5}O_3$ system both the temperature $T_{comp} = 175$K [11] and $T_{comp} = 248$K [3] are reported. There are also problems with the choice of the calculation scheme for the molecular field approximation and the incorrect interpretation of the nature of the magnetic moment compensation phenomenon, for example, as a result of the competition between single-ion spin anisotropy and the Dzyaloshinskii-Moriya interaction. In contrast to spin reorientation in weak ferromagnets with a magnetic rare-earth ion (see, for example, [1, 13, 14]), the SR transition in weak ferrimagnets $RFe_{1-x}Cr_xO_3$ with a non-magnetic R ion (La, Y, Lu) has also not yet received an adequate description.

In this paper, developing the model concepts laid down by A.M. Kadomtseva and co-workers [2], we present a consistent molecular field analysis of the magnetic properties of weak ferrimagnets of the YFe$_{1-x}$Cr$_x$O$_3$ type.

**2 Model**

The mixed systems YFe$_{1-x}$Cr$_x$O$_3$, like the "parent" orthoferrites and orthochromites, are orthorhombic perovskites with the space group *Pbnm*. There are 4 magnetic 3d ions per unit cell (see Fig. 1), for which the following classical basis vectors can be introduced:

$$4S\boldsymbol{F} = \boldsymbol{S}^{(1)} + \boldsymbol{S}^{(2)} + \boldsymbol{S}^{(3)} + \boldsymbol{S}^{(4)},$$

$$4S\boldsymbol{G} = \boldsymbol{S}^{(1)} - \boldsymbol{S}^{(2)} + \boldsymbol{S}^{(3)} - \boldsymbol{S}^{(4)},$$

$$4S\boldsymbol{C} = \boldsymbol{S}^{(1)} + \boldsymbol{S}^{(2)} - \boldsymbol{S}^{(3)} - \boldsymbol{S}^{(4)},$$

$$4S\boldsymbol{A} = \boldsymbol{S}^{(1)} - \boldsymbol{S}^{(2)} - \boldsymbol{S}^{(3)} + \boldsymbol{S}^{(4)}. \tag{1}$$

Here the vector $\boldsymbol{G}$ describes the main antiferromagnetic component of the magnetic structure (the Néel vector), $\boldsymbol{F}$ is the vector of weak ferromagnetism (overt canting of sublattices), weak antiferromagnetic components $\boldsymbol{C}$ and $\boldsymbol{A}$ describe the canting of magnetic sublattices without the formation of a total magnetic moment (hidden canting of sublattices). "Allowed" spin configurations for the $3d$ sublattice, compatible with the antiferromagnetic sign of the main isotropic superexchange, are designated as $\Gamma_1\ (A_x, G_y, C_z)$, $\Gamma_2\ (F_x, C_y, G_z)$, $\Gamma_4\ (G_x, A_y, F_z)$, where the only nonzero components of the basis vectors appear in brackets. In the process of spin reorientation, a transition from one configuration to another is possible.

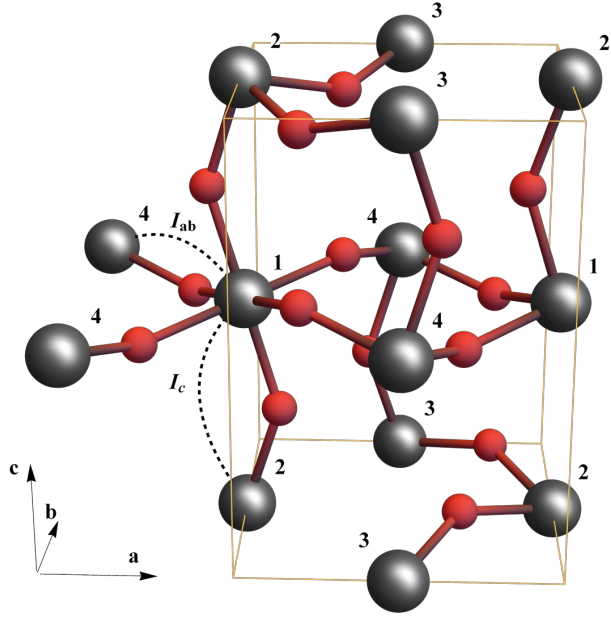

**Figure 1.** Structure of superexchange bonds; large balls are ions $Fe^{3+}$, $Cr^{3+}$, small ones are $O^{2-}$; 1, 2, 3, 4 are magnetic ions in four nonequivalent positions

Let us represent the spin-Hamiltonian of a weak ferrimagnet in its simplest form, taking into account only the contributions of the isotropic exchange interaction, the antisymmetric Dzyaloshinskii-Moriya exchange (DM), and a simplified form of the second-order single-ion spin anisotropy:

$$\widehat{H} = \widehat{H}_{ex} + \widehat{H}_{DM} + \widehat{H}_{SIA}^{(2)},$$

$$\widehat{H}_{ex} = \frac{1}{2}\sum_{\langle mn \rangle} I_{mn} (\widehat{\boldsymbol{S}}_m \cdot \widehat{\boldsymbol{S}}_n),$$

$$\widehat{H}_{DM} = \frac{1}{2}\sum_{\langle mn \rangle} \boldsymbol{d}_{mn} \cdot [\widehat{\boldsymbol{S}}_m \times \widehat{\boldsymbol{S}}_n],$$

$$\widehat{H}_{SIA}^{(2)} = D(\widehat{S}_z^2 - \frac{1}{3}S(S+1)), \tag{2}$$

where summation occurs over the nearest neighbors, $I_{mn}$ is the exchange integral, $\boldsymbol{d}_{mn}$ is the Dzyaloshinskii vector, $D$ is the anisotropy constant, which, generally speaking, differs for $Fe^{3+}$ and $Cr^{3+}$ ions (for simplicity of notation, $\widehat{H}_{SIA}^{(2)}$ summation over lattice nodes is omitted).

Figure 1 shows the structure of superexchange bonds in the model. Ions in position 1 interact with four nearest neighbors in the $ab$ plane and two along the $c$ axis. The cation-anion distances and superexchange bond angles for the nearest neighbors differ insignificantly, so below we assume the equality of the superexchange integrals $I_{ab} = I_c = I$ and modules of Dzyaloshinskii vectors, $d_{ab} = d_c = d$, although the vectors themselves are directed in different directions.

**Table 1.** Components $x$, $y$, $z$ of structure factors $[\boldsymbol{r}_m \times \boldsymbol{r}_n]$ calculated from neutron diffraction data [11] for YFe$_{0.5}$Cr$_{0.5}$O$_3$

|  | $x$ | $y$ | $z$ |
|---|---|---|---|
| $[\boldsymbol{r}_2 \times \boldsymbol{r}_1]$ | 0.216 | 0.562 | 0 |
| $[\boldsymbol{r}_4 \times \boldsymbol{r}_1]$ | ±0.303 | 0.287 | 0.397 |

As early as 1970, A.S. Moskvin obtained a microscopic expression for the relation between the Dzyaloshinskii vector and the geometry of the cation-anion-cation superexchange bond [15] (see also works [16,17] and later review articles [18-22]):

$$\boldsymbol{d}_{mn} = d_{mn}(\theta)\,[\boldsymbol{r}_m \times \boldsymbol{r}_n], \qquad (3)$$

where $\boldsymbol{r}_{m,n}$ are unit vectors along the $O^{2-}$ – $Fe^{3+}$ or $O^{2-}$ – $Cr^{3+}$ bonds, θ is the superexchange bond angle. The structural factors determining the orientation of the Dzyaloshinskii vectors in orthoferrites-orthochromites of the YFe$_{1-x}$Cr$_x$O$_3$ type are given in Table 1.

The simple formula (3) allows us to establish a direct relation between magnetic noncollinearity (overt and hidden canting of sublattices) in weak ferromagnets with a crystalline structure [16–22]. The most important result of the microscopic theory of antisymmetric exchange [17-22] was not so much the estimate of the numerical value as the prediction of the sign of the Dzyaloshinskii vector, in particular, the different sign in the pairs $Fe^{3+}$ - $Fe^{3+}$, $Cr^{3+}$ - $Cr^{3+}$ on the one hand and the pairs $Fe^{3+}$ - $Cr^{3+}$, $Cr^{3+}$ - $Fe^{3+}$ on the other hand, which played a fundamental role in the prediction and experimental discovery of a new type of magnetic ordering - weak ferrimagnetism [2,17-22].

**3 Mean field approximation**

In the simplest model of weak ferrimagnets of the type $RFe_{1-x}Cr_xO_3$ (R = La, Y, Lu), which assumes a single magnetic ordering in the Fe–Cr subsystem with molecular fields common to all $Fe^{3+}$ ($Cr^{3+}$) ions, the bilinear part of the Hamiltonian operator (2) is represented in the form

$$\widehat{H}_{ex} + \widehat{H}_{DM} = \sum_n (\boldsymbol{h}_n \cdot \widehat{\boldsymbol{S}}_n) - \frac{1}{2}\sum_n (\boldsymbol{h}_n \cdot \langle\boldsymbol{S}_n\rangle), \qquad (4)$$

where for the molecular field $\boldsymbol{h}_n$ at the node $n$, taking into account the leading contributions of the isotropic exchange and the Dzyaloshinskii-Moriya interaction, we have

$$\boldsymbol{h}_n = \sum_m \left(I_{mn}\langle\boldsymbol{S}_m\rangle + [\boldsymbol{d}_{mn} \times \langle\boldsymbol{S}_m\rangle]\right), \qquad (5)$$

here $\langle\boldsymbol{S}_m\rangle$ is the thermodynamic average spin of an arbitrary ion ($Fe^{3+}$ or $Cr^{3+}$)

$$\langle\boldsymbol{S}_m\rangle = -\frac{\boldsymbol{h}_m}{h_m} S\, B_S\left(\frac{Sh_m}{k_BT}\right), \qquad (6)$$

$B_S$ is Brillouin function, $h_m = |\boldsymbol{h}_m|$.

It is obvious that, in contrast to the homogeneous parent systems $YFeO_3$ and $YCrO_3$, for weak ferrimagnets of the $YFe_{1-x}Cr_xO_3$ type we are forced to introduce a number of additional assumptions and approximations to solve the molecular field equations (6):

1) $Fe^{3+}$ and $Cr^{3+}$ ions fill the lattice sites with equal probability;
2) the parameters of the spin Hamiltonian do not depend on either the local configuration or the concentration of $Fe^{3+}$ and $Cr^{3+}$ ions;
3) the long-range crystalline and magnetic (spin) orders are preserved, that is, the classification of possible magnetic structures ($\Gamma_{1,2,4}$) and the corresponding ratios between the average values of spin moments in positions 1, 2, 3 and 4 (see Table 2) are preserved, which makes it possible to consider the molecular field equations for only one position of the $3d$ ions.

**Table 2**. Relationship of spin components on different sublattices in phases $\Gamma_1$, $\Gamma_2$ and $\Gamma_4$

| $\Gamma_1(A_x, G_y, C_z)$ | $\Gamma_2(F_x, C_y, G_z)$ | $\Gamma_4(G_x, A_y, F_z)$ |
|---|---|---|
| $S_x^{(1)} = -S_x^{(2)} = -S_x^{(3)} = S_x^{(4)}$ | $S_x^{(1)} = S_x^{(2)} = S_x^{(3)} = S_x^{(4)}$ | $S_x^{(1)} = -S_x^{(2)} = S_x^{(3)} = -S_x^{(4)}$ |
| $S_y^{(1)} = -S_y^{(2)} = S_y^{(3)} = -S_y^{(4)}$ | $S_y^{(1)} = S_y^{(2)} = -S_y^{(3)} = -S_y^{(4)}$ | $S_y^{(1)} = -S_y^{(2)} = -S_y^{(3)} = S_y^{(4)}$ |
| $S_z^{(1)} = S_z^{(2)} = -S_z^{(3)} = -S_z^{(4)}$ | $S_z^{(1)} = -S_z^{(2)} = S_z^{(3)} = -S_z^{(4)}$ | $S_z^{(1)} = S_z^{(2)} = S_z^{(3)} = S_z^{(4)}$ |

Thus, for the molecular field $\boldsymbol{h}_{Fe}$ in position 1 we obtain

$$\boldsymbol{h}_{Fe} = P_{Fe}(x)\langle 4I_{FeFe}\,\widehat{\boldsymbol{S}}_{Fe}^{(4)} + 2I_{FeFe}\,\widehat{\boldsymbol{S}}_{Fe}^{(2)} +$$

$$+4[\boldsymbol{d}_{FeFe}^{(41)} \times \widehat{\boldsymbol{S}}_{Fe}^{(4)}] + 2[\boldsymbol{d}_{FeFe}^{(21)} \times \widehat{\boldsymbol{S}}_{Fe}^{(2)}]\rangle +$$

$$+P_{Cr}(x)\langle 4I_{FeCr}\,\widehat{\boldsymbol{S}}_{Cr}^{(4)} + 2I_{FeCr}\,\widehat{\boldsymbol{S}}_{Cr}^{(2)} +$$

$$+4[\boldsymbol{d}_{FeCr}^{(41)} \times \widehat{\boldsymbol{S}}_{Cr}^{(4)}] + 2[\boldsymbol{d}_{FeCr}^{(21)} \times \widehat{\boldsymbol{S}}_{Cr}^{(2)}]\rangle, \qquad (7)$$

where $P_{Fe}(x) = 1 - x$, $P_{Cr}(x) = x$ are the concentrations of $Fe^{3+}$ and $Cr^{3+}$ ions, respectively, brackets $\langle ... \rangle$ denote the thermodynamic average, the components of the vectors $\boldsymbol{S}^{(2)}$ and $\boldsymbol{S}^{(4)}$ are expressed through $\boldsymbol{S}^{(1)}$ in accordance with Table 2. The field $\boldsymbol{h}_{Cr}$ has the same form, but with the substitution of Fe↔Cr on the right-hand side. Note that in (7) the nonequivalent contributions of the 1-2 and 1-4 bonds are highlighted, which is especially important given the different orientation of the Dzyaloshinskii vectors for these bonds. Thus, system (6), in essence, consists of two (for the $Fe^{3+}$ ion and the $Cr^{3+}$ ion) vector self-consistent equations.

**4 Results**

The system (6) was solved numerically with the following exchange parameters: $I_{FeFe} = 36.6$ K and $I_{CrCr} = 18.7$ K, calculated from the Neel temperatures of the orthoferrite YFeO3, $T_N(0) = 640$ K [23], and orthochromite YCrO3, $T_N(1) = 140$ K [24]. The exchange integrals between the Fe - Cr and Cr - Fe ions were assumed to be the same: $I_{FeCr} = I_{CrFe} = 13.4 \pm 0.4$ K according to the data of [25]. The scalar parameters of the antisymmetric Dzyaloshinskii-Moriya exchange $d_{FeFe} = 2.0$ K and $d_{CrCr} = 1.7$ K were selected based on the data on the saturation magnetization $M = 1.5$ emu/g in the "parent" YFeO3 and YCrO3 [2]. According to the predictions of microscopic theory [17-22], the parameter $d_{FeCr} = d_{CrFe}$ should have a sign opposite to the sign of the parameters $d_{FeFe}$ and $d_{CrCr}$, its value $d_{FeCr} = d_{CrFe} = -2.5$ K was chosen to match the calculated compensation temperature with the observed $T_{comp} \approx 225$ K temperature for single-crystal YFe$_{1-x}$Cr$_x$O3 samples at $x = 0.38$ [26].

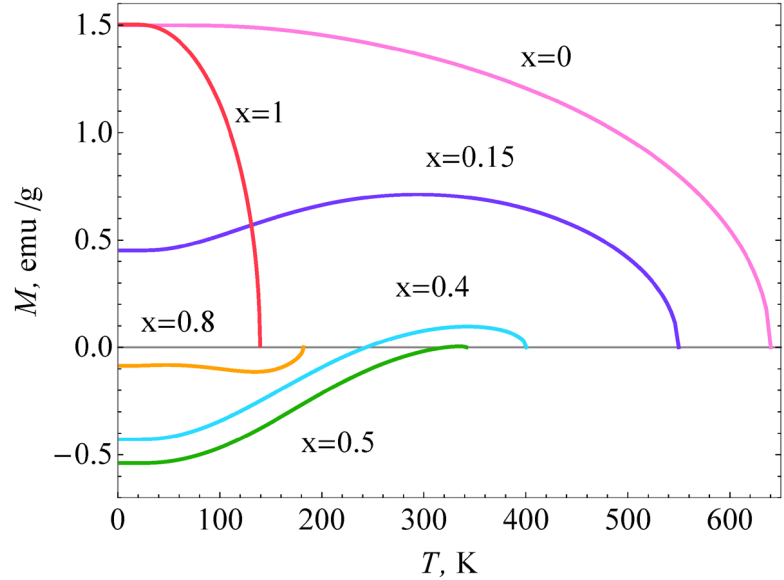

**Figure 2.** Temperature dependence of magnetization of YFe$_{1-x}$Cr$_x$O3 at different chromium concentrations $x$

Figure 2 shows the results of calculating the temperature dependences of the magnetization of the weak ferrimagnet YFe$_{1-x}$Cr$_x$O3 at some concentration values from $x = 0$ to $x = 1$ under the assumption of maintaining the magnetic configuration $\Gamma_4$. The region where $M < 0$, corresponds to negative magnetization.

It should be noted that the condition of preservation of the magnetic configuration is a critical limitation of the model: despite good qualitative agreement with the experiment, the saturation magnetization of single-crystal samples with $x = 0.38, 0.5, 0.65$ is significantly, two to three times, less than the theoretical predictions, which, in light of the transitions with a change in the orientation of the weakly ferrimagnetic moment in the $ac$ plane experimentally detected for these compositions [2], indicates the possible implementation of the spatial orientation of the antiferromagnetism vector, i.e., the configuration $G_{xyz}$.

Analysis of the model showed that when the Dzyaloshinskii vector $d_{FeCr}$ for the Fe - Cr ion pair is directed opposite to the vectors $d_{FeFe}$ and $d_{CrCr}$ (for the Fe - Fe and Cr - Cr pairs, respectively), the magnetization drops sharply with deviation from the parent compositions, but at $|d_{FeCr}| \geq |d_{FeCr}^{(cr)}|$, where $d_{FeCr}^{(cr)} \approx -1.55$ K, on the $T - x$ phase diagram a region of negative magnetization appears and grows with growth $|d_{FeCr}|$, limited by two lines of compensation points.

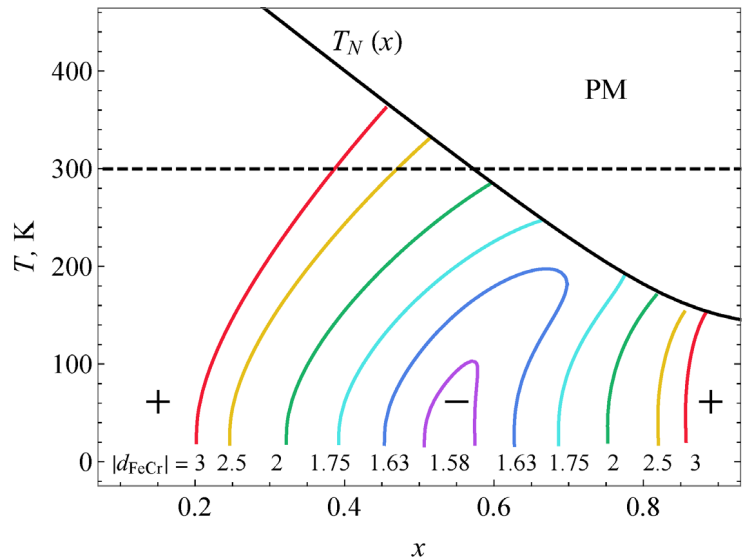

**Fig. 3.** $T - x$ phase diagram of magnetization compensation regions at different values of $d_{FeCr} < 0$. The "+" sign indicates the region of "positive" magnetization. The dotted line is room temperature $T = 300$ K

Figure 3 shows the $T - x$ phase diagram of the weak ferrimagnet $YFe_{1-x}Cr_xO_3$, where the curve $T_N(x)$ limits the region of magnetic ordering, and the curves located below indicate the lines of compensation points, i.e., the change in the sign of magnetization at different values of the parameter $d_{FeCr}$. The region between the lines of compensation points at a certain value of the parameter $d_{FeCr}$ is a region of negative magnetization. The $YFe_{1-x}Cr_xO_3$ system under consideration corresponds to two lines, or two regions, of compensation points with $|d_{FeCr}| = 2.5$, where the "wide" region is $0.25 \leq x \leq 0.5$ and the "narrow" region near $x \approx 0.83$.

We note the existence of a fairly large range of parameter values $d_{FeCr}$ at which the compensation temperature in the weak ferrimagnet $YFe_{1-x}Cr_xO_3$ falls within the practically most important range of room temperatures of the order of $T = 300$ K, in particular, with the parameter we have chosen $d_{FeCr} = d_{CrFe} = -2.5$ K compensation temperature reaches room temperature $T = 300$ K in composition with $x \approx 0.45$.

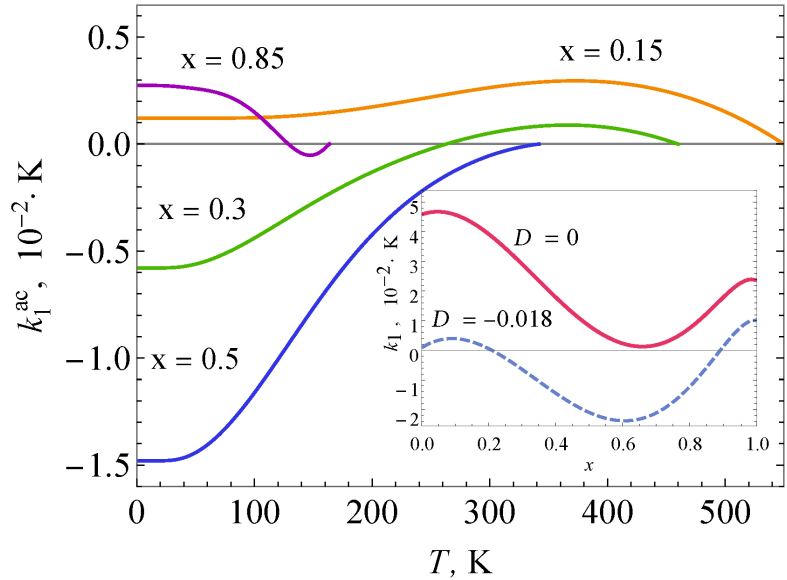

**Figure 4.** Example of temperature dependence of the first anisotropy constant $k_1$ for $ac$-plane with single-ion anisotropy $D = -0.018$ K and at different chromium concentrations $x$; the inset shows the dependence $k_1(x, T = 0)$

Unlike $YFeO_3$ and $YCrO_3$, which are weak ferromagnets with a basic magnetic structure of the type $\Gamma_4$ below the Neel temperature, weak ferrimagnets orthoferrites-orthochromites YFe$_{1-x}$Cr$_x$O$_3$ exhibit complete or partial spin-reorientation of the type $\Gamma_4 - \Gamma_2$ in a wide range of substitution [2]. Such unexpected behavior, usually typical for orthoferrites with magnetic rare-earth ions (Er, Tm, Dy, ...), is explained mainly by a strong decrease in the contribution of the DM interaction to the magnetic anisotropy. This is evident when comparing the free energy $\Phi_{\Gamma_4}$ in phase $\Gamma_4$ with the energy $\Phi_{\Gamma_2}$ in phase $\Gamma_2$. Calculations show (see Fig. 4) that the first anisotropy constant $k_1 = \frac{1}{2}(\Phi_{\Gamma_2} - \Phi_{\Gamma_4})$ experiences a minimum near the concentration $x \approx 0.65$, i.e. here the phase $\Gamma_4$ is less favorable, in comparison with the parent compositions. When taking into account single-ion anisotropy $\widehat{H}_{SIA}^{(2)}$, the constant $k_1$ can become negative, which explains the transition to the phase $\Gamma_2$.

## 5 Conclusion

We have considered the spin-Hamiltonian of the YFe$_{1-x}$Cr$_x$O$_3$ system taking into account the main isotropic and anisotropic interactions. Within the framework of the molecular field approximation, the average value of the magnetic moments of the $3d$ ions and the effective anisotropy constant have been calculated. The existence of two regions of compensation points in the YFe$_{1-x}$Cr$_x$O$_3$ model system, "wide" $0.25 \leq x \leq 0.5$ and "narrow" near $x \approx 0.83$, limiting the region of negative magnetization has been shown. For the system under consideration, the compensation temperature reaches room temperature at $x \approx 0.45$. The phenomenon of spin reorientation observed for single-crystal samples in a wide concentration range is explained by a sharp decrease in the contribution of the antisymmetric exchange to the magnetic anisotropy with increasing deviation from the parent compositions and competition with the contribution of the single-ion anisotropy of the Fe and Cr ions. It is suggested that the spatial orientation of the antiferromagnetism vector (Néel vector) and configuration $G_{xyz}$ are the cause of the small value of saturation magnetization observed experimentally for compositions inside or near the region of negative magnetization.

**Financing the work**

The work was supported by the Ministry of Science and Higher Education of the Russian Federation, project FEUZ-2023-0017**References**

[1] K.P. Belov, A.K. Zvezdin, A.M. Kadomtseva, R.Z. Levitin. Orientational transitions in rare-earth magnets. Nauka, Moscow (1979). 317 p.

[2] A.M. Kadomtseva, A.S. Moskvin, I.G. Bostrom, B.M. Wanklin, N.A. Khafizova. JETP **72**, 2286–2297 (1977).

[3] Mao Jinhua et al. App. Phys. Lett. **98**, 192510 (2011).

[4] N. Dasari, P. Mandal, A. Sundaresan, N. S. Vidhyadhiraja. Europhys. Lett. **99**, 17008 (2012).

[5] T. Bora, S. Ravi. J. Appl. Phys. **114**, 033906 (2013).

[6] V. Nair, V. Subramanian, P. Santhosh. Journal of Applied Physics **113**, *21* (2013).

[7] F. Pomiro, R. D. Sánchez, G. Cuello, A. Maignan, C. Martin, and R. E. Carbonio. Phys. Rev. B **94**, 134402 (2016).

[ 8] O.V. Billoni, F. Pomiro, S.A. Cannas, C. Martin, A. Maignan, R.E. Carbonio, J. Phys.: Condens. Matter **28**, 476003 (2016).

[9] APG Rodrigues, MA Morales, RB Silva, DRAB Lima, RLBA Medeiros, JH Ara´ujo, DMA Melo. Journal of Physics and Chemistry of Solids **141**, 109334 (2020).

[10] R. Salazar-Rodriguez, D. Aliaga Guerra, J.-M. Greneche, K. M. Taddei, N.-R. Checca-Huaman, E. C. Passamani, J. A. Ramos-Guivar. Nanomaterials **12**, *19*, 3516 (2022).

[11] J. Yang, H. Cao, Z. Lu, J. Mo, Y. Zhou, K. Gao, Y. Xia, M. Liu. Physica Status Solidi B **260**, *7*, 2300145 (2023).12